\documentclass{PoS}

\title{High Momentum Probes of Nuclear Matter}

\ShortTitle{High Momentum Probes of Nuclear Matter}

\author{\speaker{Rainer Fries} \\
        Cyclotron Institute, Texas A\&M University, USA
        and RIKEN/BNL Research Center, USA\\
        E-mail: \email{rjfries@comp.tamu.edu}}

\abstract{We discuss how the chemical composition of QCD jets
is altered by final state interactions in surrounding nuclear matter.
We describe this process through conversions of leading jet particles.
We find that conversions lead to an enhancement of kaons at high 
transverse momentum in Au+Au collisions at RHIC, while their azimuthal 
asymmetry $v_2$ is suppressed.}

\FullConference{4th international workshop High-pT physics at LHC 09\\
                 February 4-7, 2009\\
                 Prague, Czech Republic}

\begin{document}

Jets of high energy particles can be used as probes to study dense and hot 
phases of nuclear matter \cite{Wang:1991xy,BDMPS:96,Zakharov:96,
Wiedemann:2000tf,gyulassy,AMY:02}. Experiments at the 
Relativistic Heavy Ion Collider (RHIC) have shown that quark gluon plasma 
just above the phase transition temperature is a very opaque substance.
These measurements usually compare single particle spectra and two-particle 
correlations at intermediate and high transverse momenta $p_T$ in nuclear
collisions to those in elementary $p+p$ collisions. The quantitative 
outcome of these studies are estimates for the transport coefficient 
$\hat q =\mu^2/\lambda$ in hot nuclear matter, the average momentum transfer 
squared per mean free path.
Recently, it has been suggested that studying the yield of particles
at high $p_T$ can yield much more information by adding the flavor of hadrons 
as an additional observable \cite{weiliu}.

Obviously, fast quarks and gluons passing through nuclear matter can
change identity through conversion processes. In other words, the ``flavor''
(loosely used as a notation for the identity of a particle) of a leading jet
parton is not conserved. This has a variety of observable consequences.
Let us first point out that the rate at which conversions
between different flavors occur depends mainly on the total cross section of
a conversion channel. It is therefore sensitive to the mean free path $\lambda$
of a particle, and the chemical composition of the medium, rather than to the 
transport coefficient $\hat q$.

It was first pointed out
in 2002 that light quarks and gluons passing through quark gluon plasma
can ``convert'' into real and virtual photons through Compton and
annihilation processes with thermal partons, $q+\bar q \to \gamma +g$ and 
$g+q\to \gamma+q$, resp.\ \cite{fries1}. Here, we use a notation where 
the momentum of the first particle mentioned on both sides of the reaction 
equation is much larger than the momentum of the second particle in the 
lab frame and it is therefore considered to be either the leading
parton of a jet or another high-$p_T$ particle.
This source of photons has been studied by several authors since
\cite{FMS:05,SGF:02,GAFS:04,simon,Turbide:2007mi}. It appears to
make a non-negligible contribution to the total direct photon spectrum
for transverse momenta of several GeV/$c$.

\begin{figure}[b]
\centerline{
\includegraphics[width=6.0cm,angle=-90]{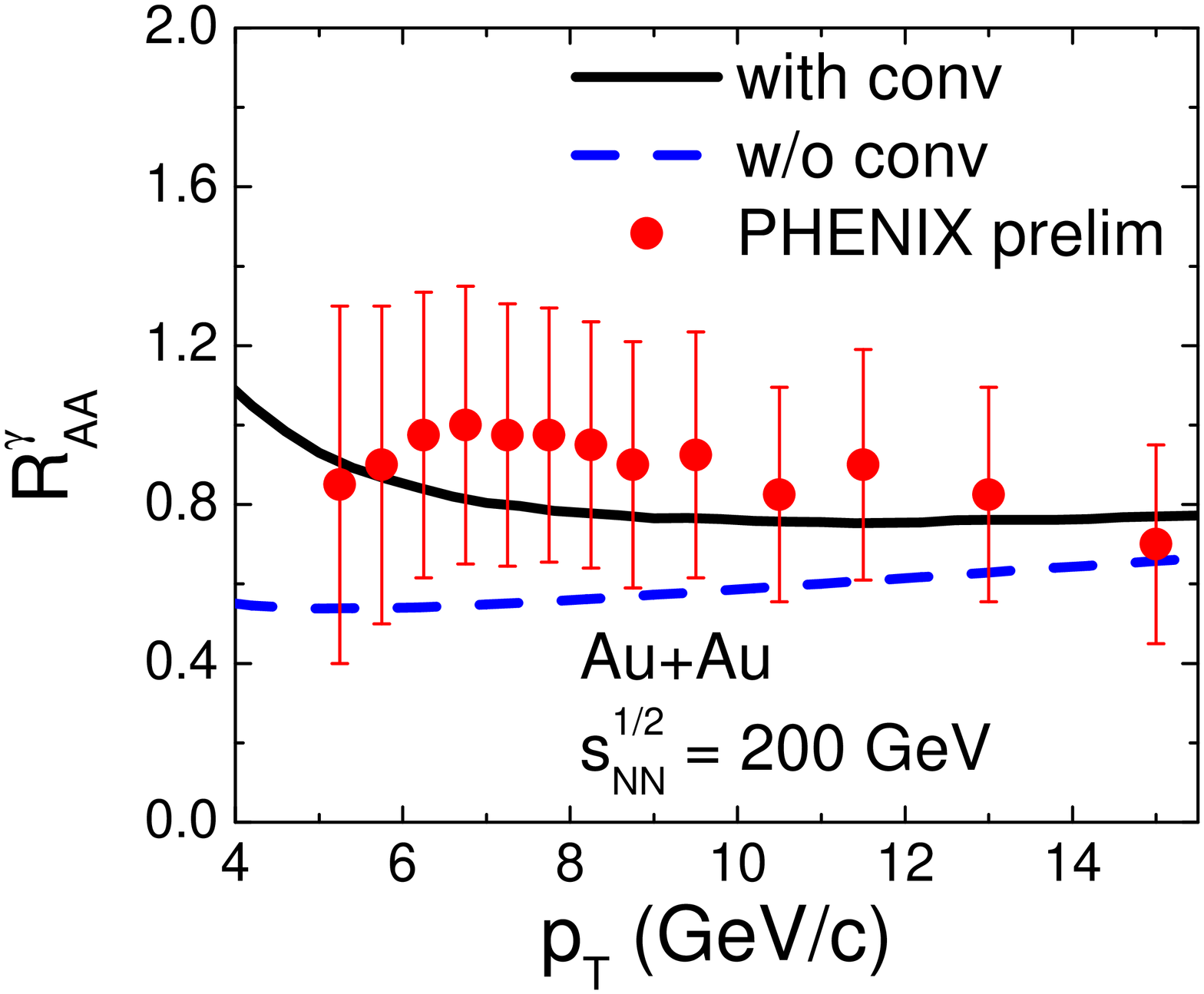}
%\hspace{-0.2cm}
\includegraphics[width=6.0cm,angle=-90]{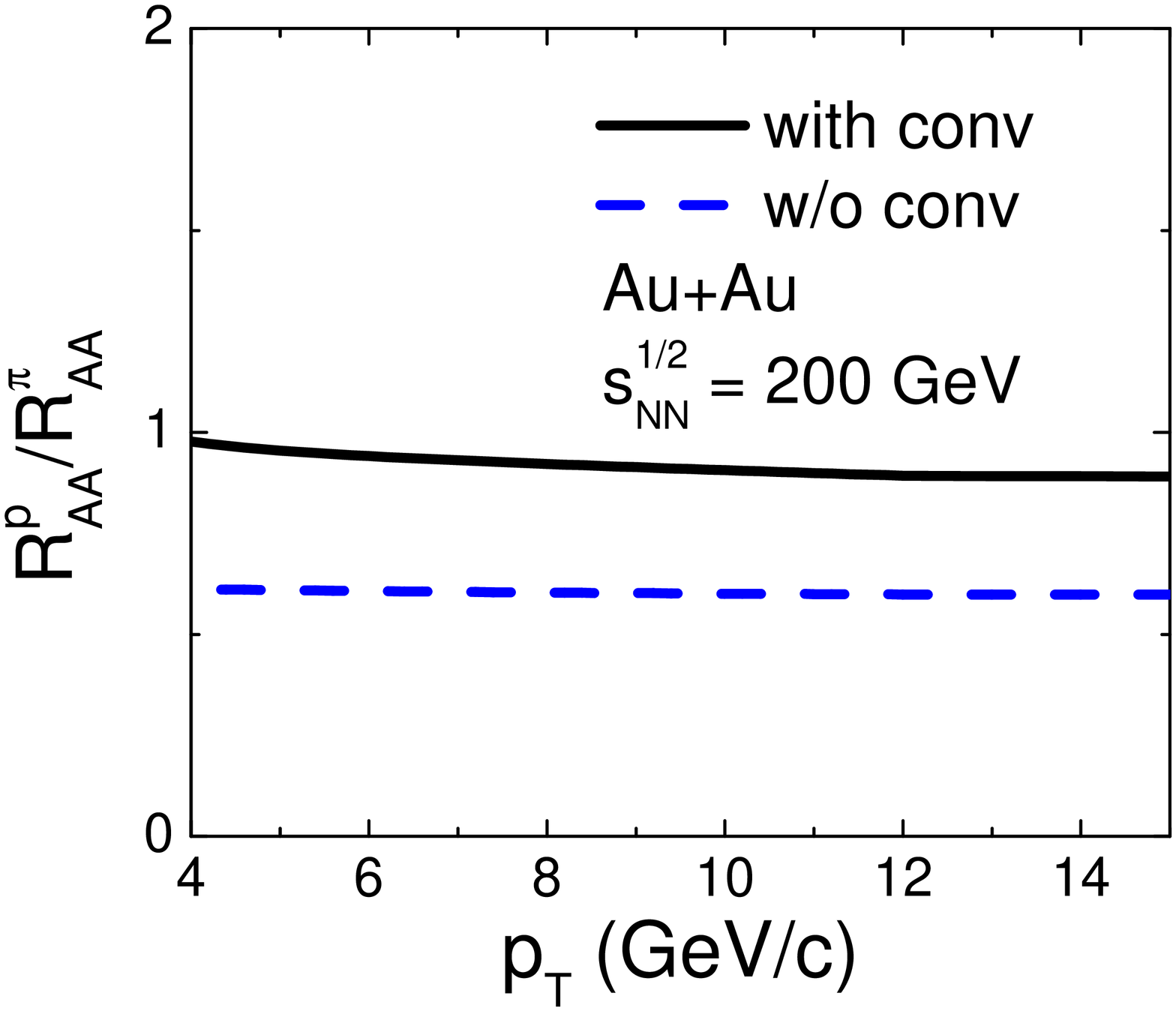}}
\caption{Left panel: The nuclear modification factor $R_{AA}$ for direct 
photons with and without conversions switched on, calculated in the model 
introduced in \cite{weiliu} (preliminary PHENIX data from \cite{Adler:2005ig}).
Right panel: The ratio of nuclear modification
factors for protons and pions is approaching one if conversions are allowed.}
\label{fig:1}
\end{figure}

Flavor changing channels and conversions for gluons into quarks and vice 
versa, $q+g \leftrightarrow g+q$ and $q+\bar q \leftrightarrow g+g$,
have been discussed by several authors \cite{weiliu1,cmk,Schafer:2007xh} in
the past. 
Recent interest in this process was fueled by the idea that partonic 
energy loss predicts a color factor 9/4 for the relative quenching
strength of gluons and quarks.
This could in principle translate into an observable difference in
quenching between various hadron species, depending on the branching ratios
of partons into hadrons via the fragmentation process. A primary candidate 
to look for this signature is the proton over pion ratio. In some 
state-of-the-art fragmentation functions like the AKK set \cite{akk} 
protons have a 
large contribution from gluon fragmentation, leading to larger suppression 
of protons in this picture due to stronger gluon quenching. No such 
suppression was found by the STAR experiment \cite{Abelev:2006jr}. A lack of 
suppression would still be consistent
with a partonic origin of energy loss if conversions are taken into account.
A typical high-$p_T$ particle on its way out of the fireball could change
identity several times between quark and gluon, leading to a rapid fading
of the difference in color factors for any observable.
As a word of caution, it it clear that the interpretation of this particular
observable has a large uncertainty due to the input from fragmentation
functions.

Examples for the nuclear modification factor of photon (left) and the ratio
of proton and pion nuclear modification factors $R_{AA}$ (right) can be found
in Fig.\ \ref{fig:1}. Here, conversions were modeled 
by propagating leading jet partons through a fireball simulation with elastic 
conversion processes described by rate equations. The details of this 
calculation can be found in Ref.\ \cite{weiliu}. The $K$ factor for the
elastic parton cross sections are kept variable to illuminate different
scenarios: $K=0$ (no conversions), $K=1$ (conversions dominated by elastic 
processes), and $K=4$ (mimicking conversions subject to much stronger coupling
than given by elastic scattering).
It can be seen from Fig.\ \ref{fig:1} that conversions into photons are 
consistent with the observed photon yield. Conversions between quarks 
and gluons with large $K$ factor are needed to avoid a larger relative
suppression of protons.

\begin{figure}[t]
\centerline{
\includegraphics[width=8.0cm]{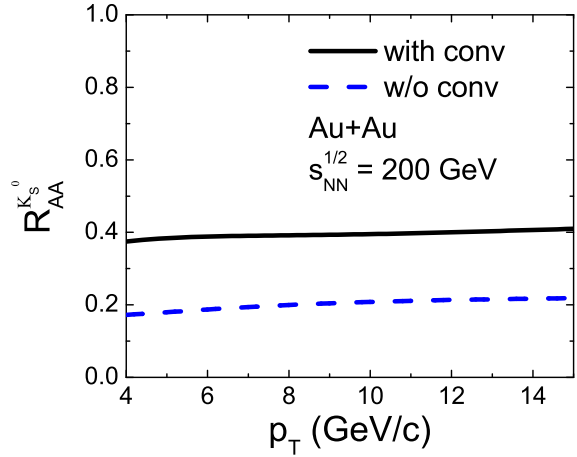}}
%\hspace{-0.2cm}
%\includegraphics[width=3.0in,height=3.0in,angle=-90]{photon_raa.eps}}
\caption{$R_{AA}$ for neutral kaons with and without conversion processes
allowed. The strangeness in the jet sample is driven towards equilibrium
by coupling it chemically to the quark gluon plasma.}
\label{fig:2}
\end{figure}

It was pointed out in Ref.\ \cite{weiliu} that strangeness at RHIC energies
could be the perfect observable to measure conversions of leading jet partons.
The bigger picture here is as follows. It is reasonable to believe
that strange quarks are chemically equilibrated in the quark gluon plasma
at RHIC, with a ratio of strange to light quarks, $w_{\rm med} = s/(u+d)
 \approx 0.5$.
On the other hand, at RHIC energies, initial jet production at intermediate
and large momenta proceeds mostly through Compton scattering of quarks
out of the initial parton distributions, hence suppressing strange quarks
in the initial jet sample, $w_{\rm jet} < 0.1$. This large chemical
imbalance in the jet sample has an opportunity to equilibrate through
their coupling to the quark gluon plasma. Indeed, the ratio of strange 
to light quarks in the jet
sample increases significantly if conversions are allowed. This translates
naturally into an enhancement of kaons over pions --- or actually into
less suppression of kaons vs pions in the nuclear modification factor
$R_{AA}$. Fig.\ \ref{fig:2} shows the change in $R_{AA}$ for neutral
kaons expected with conversions.

\begin{figure}[t]
\centerline{
\includegraphics[width=6.0cm,angle=-90]{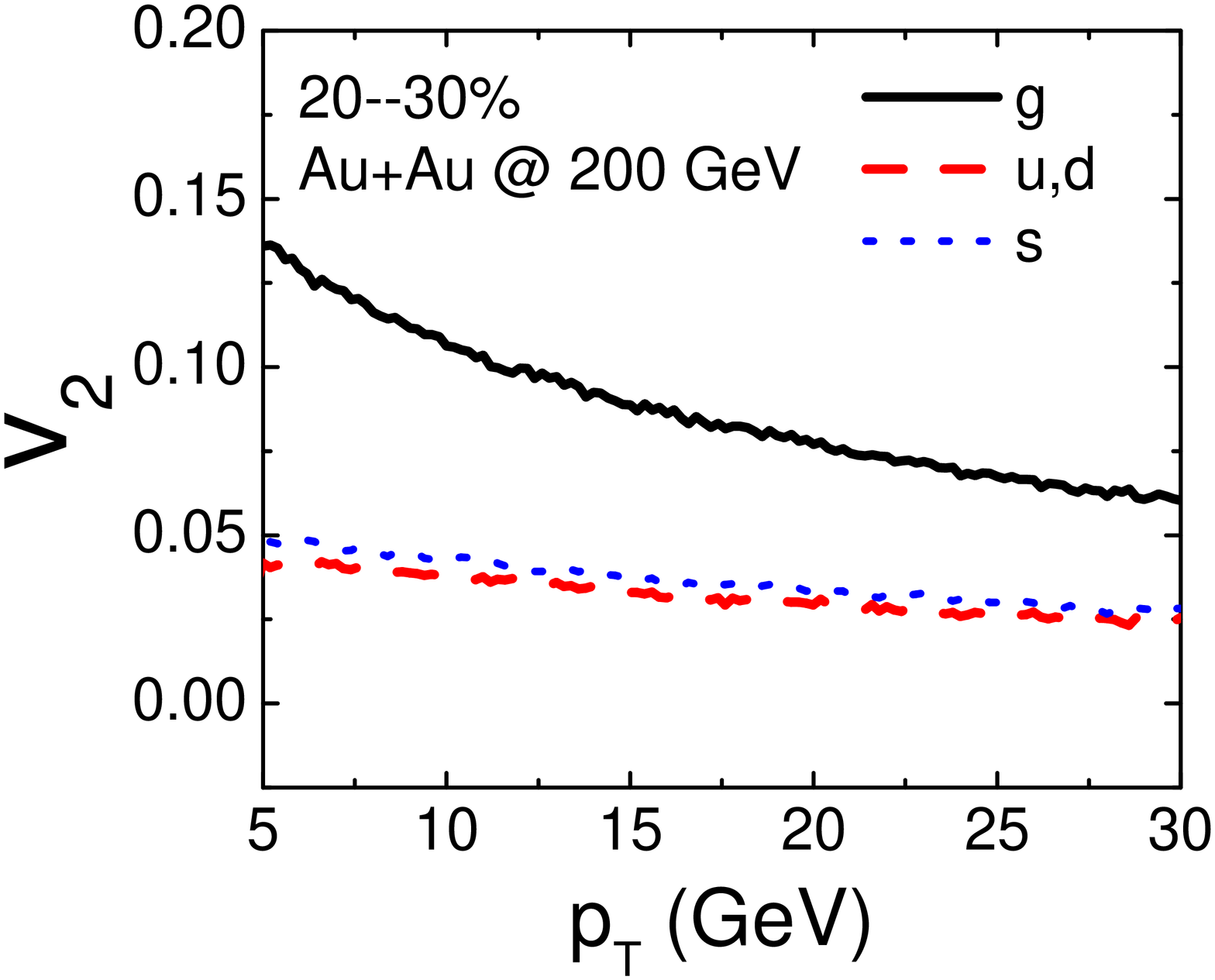}
%\hspace{-0.2cm}
\includegraphics[width=6.0cm,angle=-90]{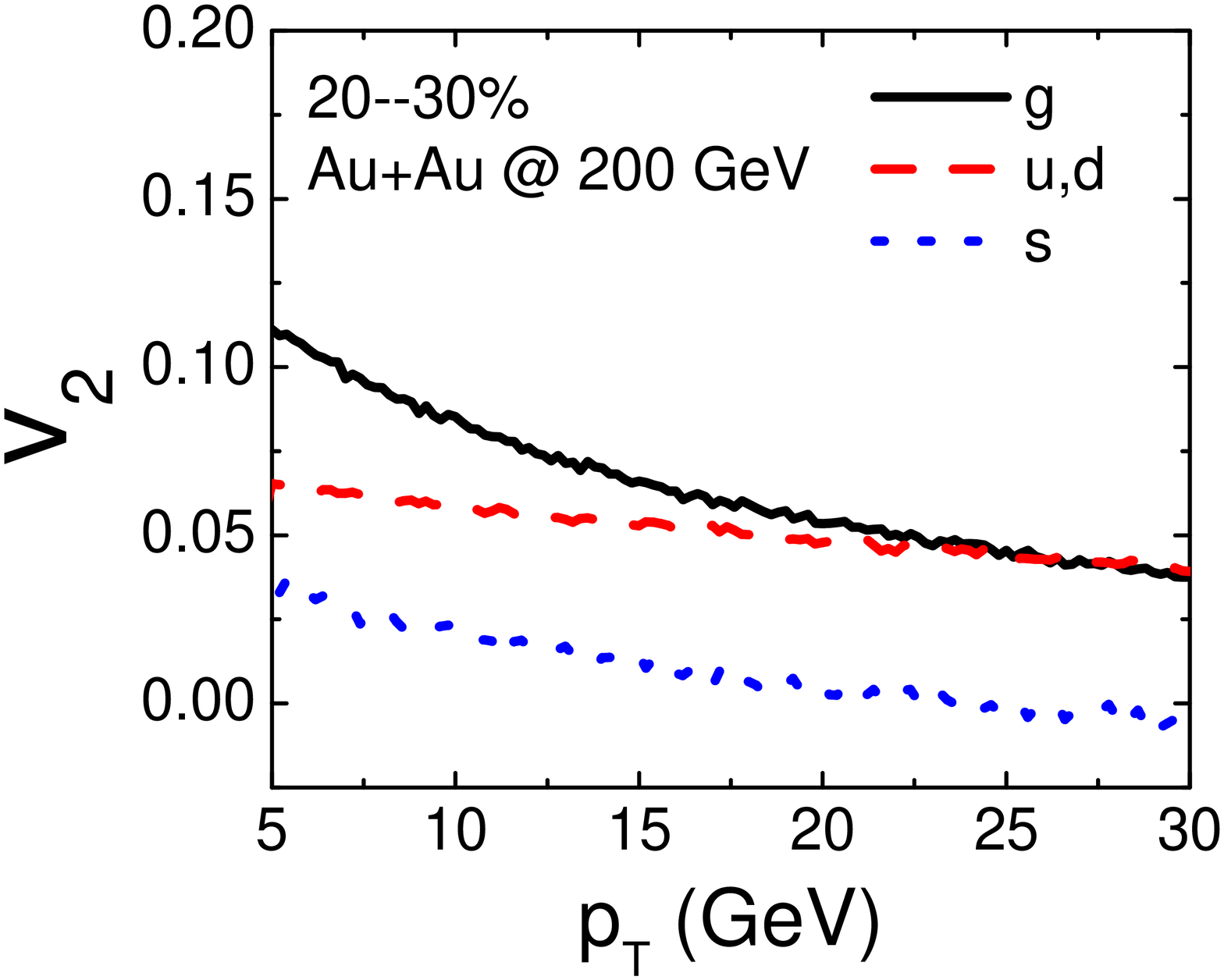}}
\caption{Left panel: The azimuthal asymmetry $v_2$ for light quarks, strange
quarks and gluons without conversions. Right panel: the same with conversions.
The $v_2$ for light quarks and gluons is not similar, while strange quarks
exhibit a suppression.}
\label{fig:3}
\end{figure}

This mechanism is obscured at LHC energy where the initial jet sample
is already close to equilibrium due to the dominance of the $g+g$
fusion channel in initial jet production. It also turns out that
heavy quarks can not play the same role as conversion signatures at LHC 
that strangeness holds at RHIC. The obstacle is that charm quarks will not 
equilibrate
chemically in the quark gluon plasma at LHC. With both low- and high-$p_T$ 
charm quarks produced perturbatively in initial hard scatterings
the chemical gradient is not large enough to see a noticable enhancement
at high $p_T$ \cite{Liu:2008bw}.

\begin{figure}[t]
\centerline{
\includegraphics[width=7.0cm,angle=-90]{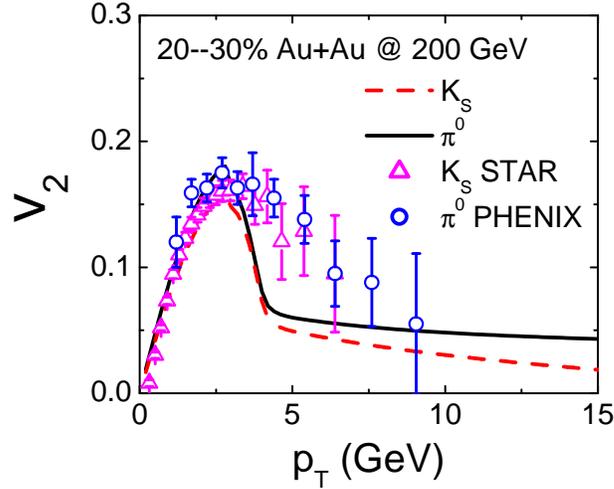}}
%\hspace{-0.2cm}
%\includegraphics[width=4.0cm,angle=-90]{quark_v2.eps}}
\caption{The resulting azimuthal asymmetry $v_2$ for kaons which is 
expected to be suppressed compared to that of pions. Data from
\cite{david,abelev}.}
\label{fig:4}
\end{figure}

Yet another signature for conversions of high $p_T$ particles in the medium
should be seen in their azimuthal asymmetry $v_2$. It had first been realized
for photons \cite{Turbide:2005bz} that conversions lead to a higher conversion 
yield
on the thicker side of the medium. This is opposite to the suppression driven
$v_2$ which would lead to less yield on the thicker side of the medium.
Hence a conversion driven azimuthal asymmetry should lead to negative $v_2$.
Of course, all sources of a given particle have to be added up for the total
yield, leading to cancellations in $v_2$. For direct photons it was forecast
that the cancellation is rather complete and that $v_2$ should
be numerically very small \cite{Turbide:2005bz,weiliu,Chatterjee:2005de,Kopeliovich:2007sd}. This is in accordance with
first measurements of direct photon elliptic flow.

We also find that the conversion mechanism leads to a reversed azimuthal 
asymmetry for additional strange quarks \cite{Liu:2008kj} at high $p_T$
at RHIC energies. After adding the effect of 
energy loss the total $v_2$ stays positive but is still significantly 
suppressed compared to light quarks and gluons. This is shown in Fig.\ 
\ref{fig:3}. Without conversions, light and strange quarks exhibit
the same azimuthal asymmetry, while gluons show larger values due to the larger
color factor (left). With 
conversions enabled, light quarks and gluons exhibit comparable $v_2$ while 
the asymmetry for
strange quarks is suppressed (right). Details of the calculation can be found in
\cite{Liu:2008kj}. Finally, Fig.\ \ref{fig:4} shows the effects translated into the
$v_2$ for neutral kaons. It is expected that their $v_2$ shows a noticable
suppression starting at about 6 to 7 GeV/$c$ compared with pions.

To summarize, we have argued that conversions of leading jet particles
is a natural consequence of their interactions with a surrounding medium.
In particular, we expect additional photons and strange hadrons at
RHIC energies. While the new photon source will increase in relative
brightness at LHC we predict the reduced kaon suppression to dissappear
at the larger collider energy. $v_2$ suppression at large $p_T$ 
for both photons and kaons could provide additional insight. 
Measuring effects from conversion processes could provide complementary
information about the mean free path of fast partons in quark gluon plasma.

RJF would like to thank the organizers of the Workshop \emph{High-$p_T$ 
physics at LHC} for their kind invitation.

\end{document}